\documentclass[a4paper]{jpconf}
\usepackage{graphicx}
\begin{document}

\title{Measurement of differential cross sections of single top-quark production in the $t$ channel in proton-proton collisions at $\sqrt{s} = 8$~TeV}
\author{Steffen R\"ocker on behalf of the CMS Collaboration}

\address{Institut f\"ur Experimentelle Kernphysik, Karlsruhe Institute of Technology, Wolfgang-Gaede-Str. 1, 76131 Karlsruhe, Germany}

\ead{steffen.roecker@kit.edu}

\begin{abstract}
Differential measurements of single top quark $t$-channel cross sections as a function of
the transverse momentum and the absolute value of the rapidity of the top quark are presented.
The data collected by the CMS experiment at the LHC at a center-of-mass energy of 8~TeV corresponds to an
integrated luminosity of 19.7 fb$^{-1}$.
Leptonic decay channels of the top quark, with either a muon or an electron in the final state, are considered.
Neural networks are used to separate signal from background contributions.
After correcting for selection efficiencies and detector resolution with an unfolding technique, the resulting distributions are
found to agree with predictions from different Monte Carlo generators within the estimated uncertainties.
\end{abstract}

\section{Introduction}

Electroweak production of single top quarks is now well established at the Tevatron~\cite{Aaltonen:2009jj,Abazov:2009ii} and at the LHC~\cite{Khachatryan:2014iya,Aad:2014fwa}. Enough data has been collected at the LHC to perform differential measurements, which are well suited to estimate the accuracy of various Monte Carlo (MC) generators. The calculations for single top $t$-channel production can either be performed in the 4 flavor scheme (4FS) or 5 flavor scheme (5FS). In the 5FS, the b quark is treated as a massless quark and is included in the parton density functions (PDFs). In the 4FS, the calculations are supposed to be more realistic as the b quark is treated as an independent massive particle. However, here the soft non-pertubative component seems to be not correctly described . These different approaches have been studied and found to have differences in the modeling of the top-quark kinematic distributions~\cite{Campbell:2009ss}. A third method used by other MC generators, e.g. \textsc{CompHEP}, combines the $2\to2$ and $2\to3$ LO diagrams in order to get a matched prediction. In the following, a recently published analysis~\cite{TOP-14-004} measuring differential cross sections to study these different approaches will be described.

\section{Differential cross section measurement}

\subsection{Data and simulation}

The full dataset of 19.7 fb$^{-1}$ recorded by the CMS experiment~\cite{CMS} at 8~TeV is used for the analysis.
The signal process is simulated with \textsc{PowHeg} interfaced to \textsc{Pythia6}. The most important backgrounds, the production of top quark anti-quark pairs ($t\bar{t}$) and of W bosons in association with jets (W+jets), as well as Z+jets production, are modeled with \textsc{MadGraph} interfaced to \textsc{Pythia6}. Diboson production (WW, WZ, ZZ) is simulated with \textsc{Pythia6}. QCD multijet production is estimated from an anti-isolated sideband region in data.

\subsection{Event selection and reconstruction}

The event selection requires events to contain exactly one isolated muon or electron, one b-tagged jet, one untagged jet and a significant amount of missing transverse energy (electron channel) or transverse W boson mass (muon channel). Events with an additional soft lepton are vetoed. Events stemming from tau decay to muons or electrons are also selected when satisfying the respective charged lepton requirements. The assignment of the selected jets is straightforward: the tagged jet is assigned to the b quark from the top quark decay and the untagged jet is assigned to the light quark in the final state. The W boson is reconstructed from the charged lepton and the neutrino vectors. The transverse components of the latter are taken from missing transverse energy and the z component of the neutrino momentum is derived from a W boson mass constraint.
\\
In addition to the beforehand defined signal region, two control regions are defined. The $t\bar{t}$ production region consists of three jets of which two have to be b-tagged (3j2t) and is used to check the modeling of the $t\bar{t}$ process and to reduce its uncertainty in the background estimation. The W+jets control region with two non-tagged jets (2j0t) is solely used to check the modeling of the W+jets process.

\subsection{Event classification and background estimation}

\begin{figure}[htb]
\begin{minipage}{18pc}
\includegraphics[width=18pc]{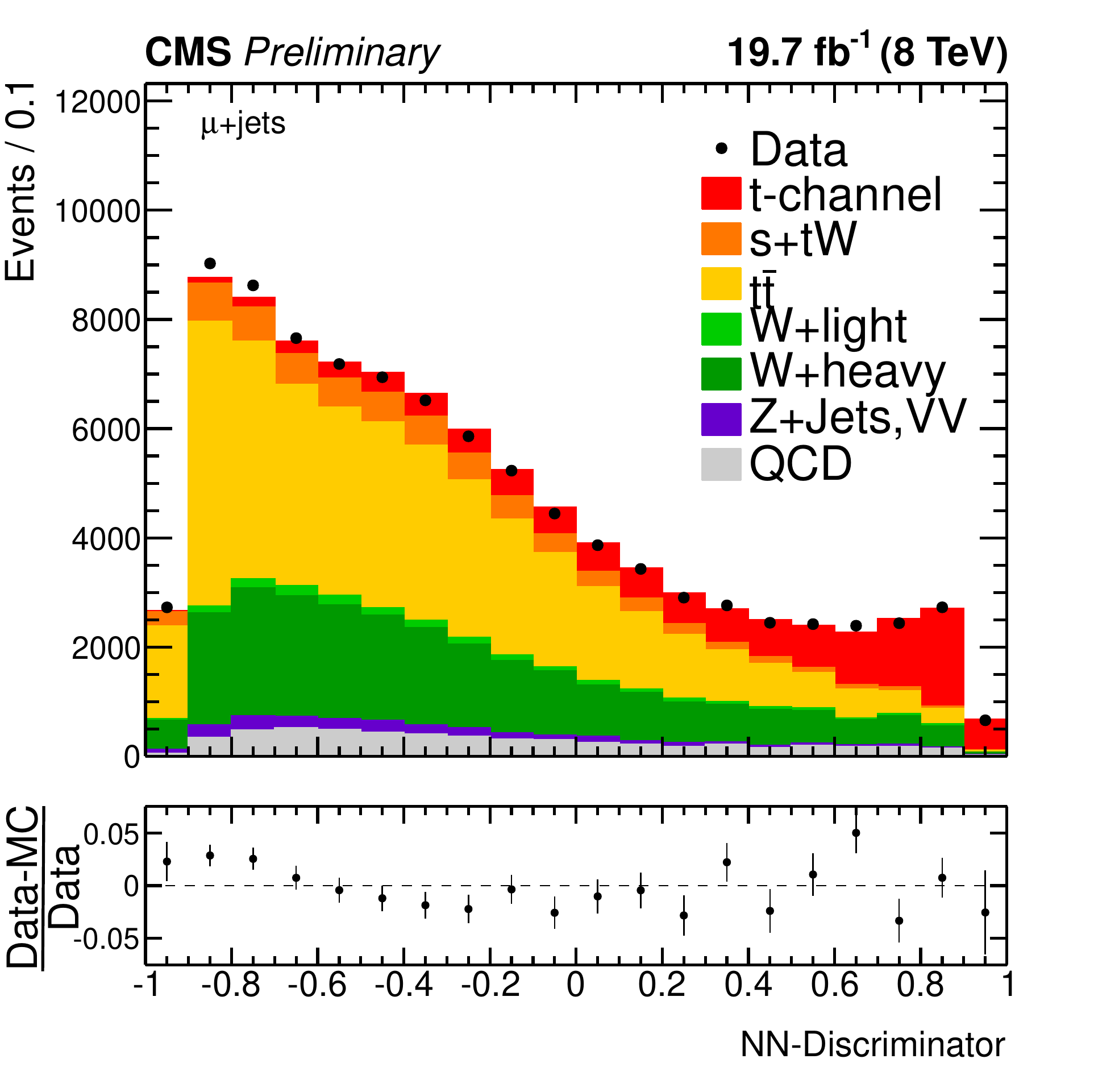}
\caption{\label{nnmu2j1t}NN discriminator output of the signal region (2j1t) in the muon channel. The templates of the different processes are normalized to the fit result.}
\end{minipage}\hspace{2pc}%
\begin{minipage}{18pc}
\includegraphics[width=18pc]{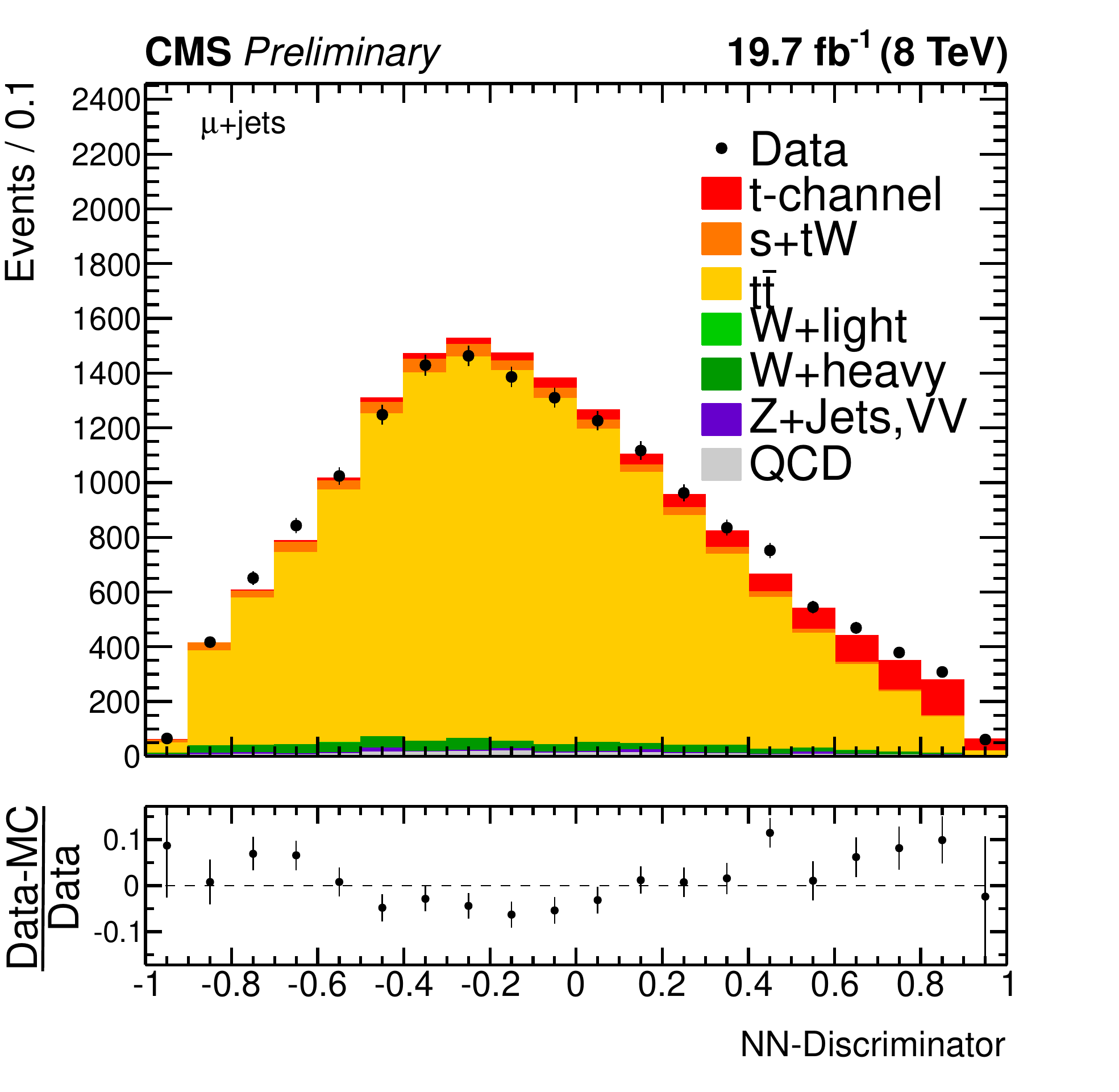}
\caption{\label{nnmu3j2t}NN discriminator output of the $t\bar{t}$ control region (3j2t) in the muon channel. The templates of the different processes are normalized to the fit result.}
\end{minipage}\hspace{2pc}%
\end{figure}

A neural network (NN) is trained in each lepton channel in order to further separate signal from backgrounds. The training of the network is performed in the signal region with 18 input variables in each channel and applied to the signal, $t\bar{t}$ and W+jets control region. Fig.~\ref{nnmu2j1t} and Fig.~\ref{nnmu3j2t} exemplarily show the NN discriminator in the muon channel 2j1t and 3j2t regions.

\begin{figure}[htb]
\begin{minipage}{18pc}
\includegraphics[width=18pc]{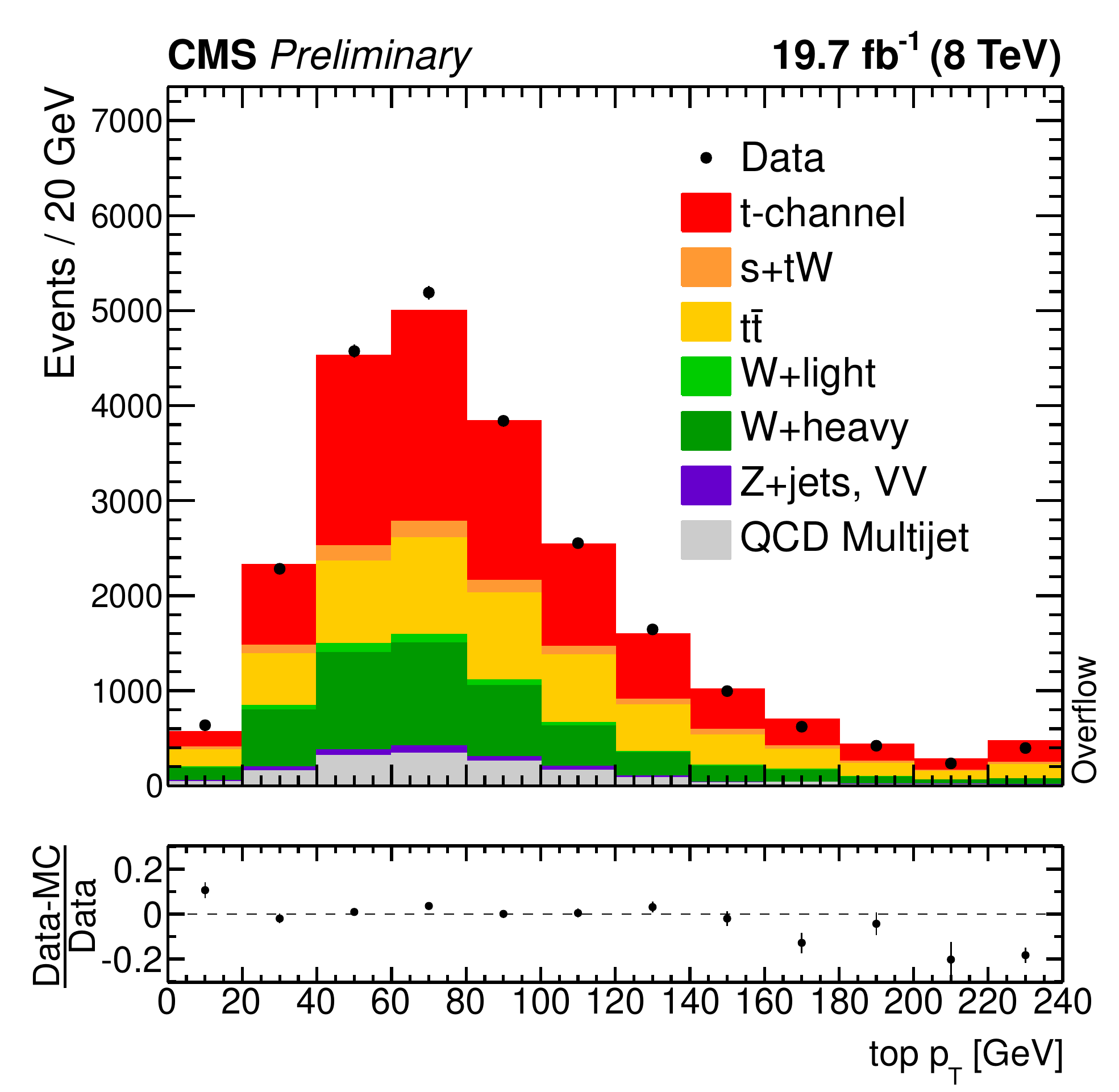}
\caption{\label{variable_top_pt}Distribution of the reconstructed top quark $p_{\mathrm{T}}$ in the combined lepton+jets channel with applied cut on the NN discriminator. The templates of the different processes are normalized to the fit result.}
\end{minipage}\hspace{2pc}%
\begin{minipage}{18pc}
\includegraphics[width=18pc]{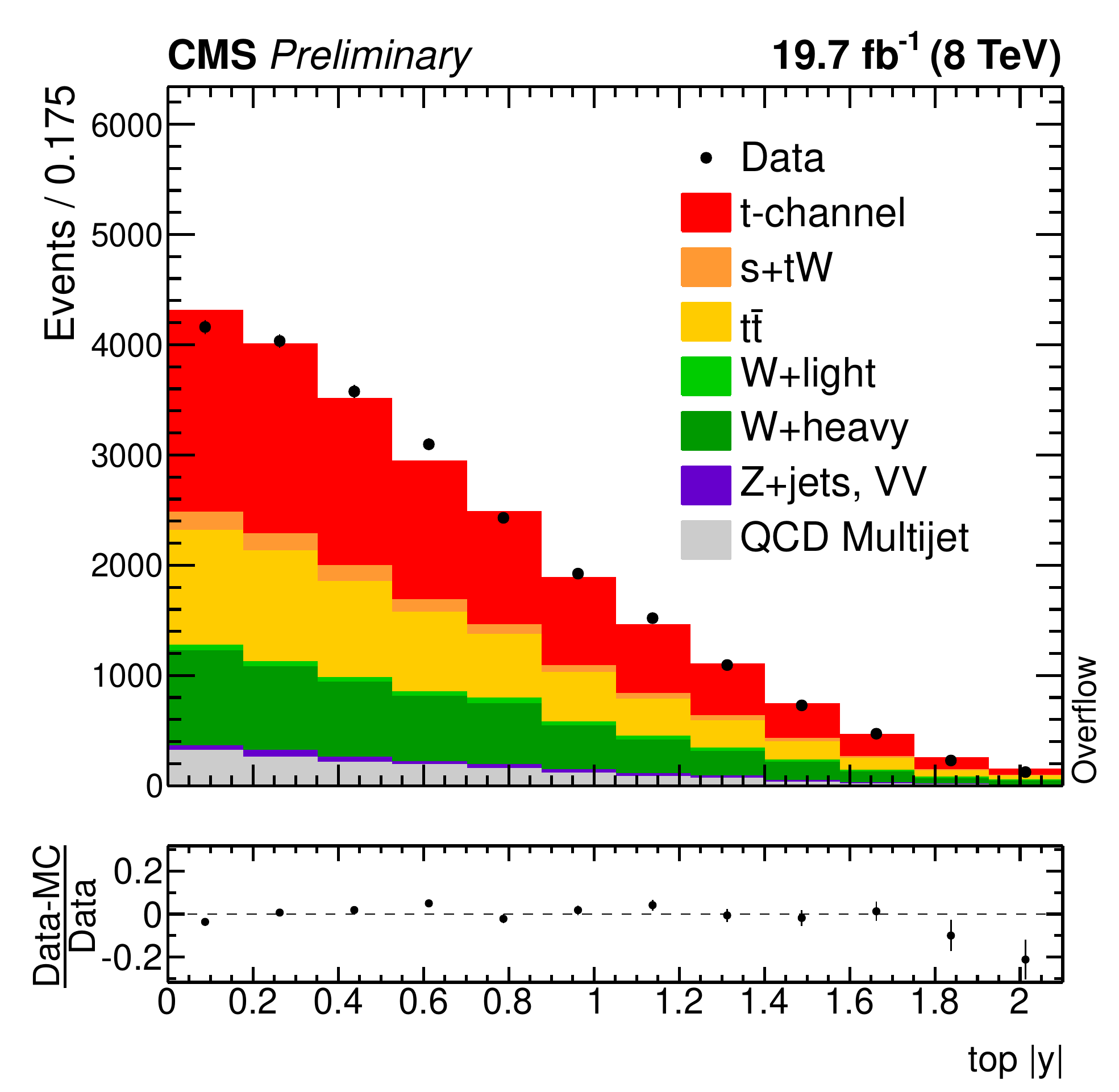}
\caption{\label{variable_top_y}Distribution of the reconstructed absolute value of the top quark rapidity $|y|$ in the combined lepton+jets channel with applied cut on the NN discriminator. The templates of the different processes are normalized to the fit result.}
\end{minipage} 
\end{figure}

The amount of background events is estimated with a simultaneous binned likelihood fit to the neural network discriminator in the signal region (2j1t) and $t\bar{t}$ enriched region (3j2t), separately for each lepton channel. In order to get an enriched signal sample, a cut is applied on the signal region discriminator output and the resulting distributions for transverse top quark momentum and absolute rapidity in both lepton channels are combined. The resulting distributions, shown in Fig.~\ref{variable_top_pt} and Fig.~\ref{variable_top_y}, are used to estimate the differential cross sections.

\subsection{Detector resolution and efficiency corrections}

The reconstructed distributions are distorted with respect to the predicted parton-level distributions due to the limited detector resolution, non-constant selection efficiency and ambiguities in the assignment of reconstructed objects to the signal event topology. After subtracting the background processes, a regularized unfolding technique utilizing \textsc{TUnfold}~\cite{TUnfold} is used to correct these effects. The regularization parameter is estimated by minimizing the global bin-by-bin correlation. Several thousand pseudo experiments are performed as a closure test, showing negligible bias and deviation.
The migration matrices, which give the probability of an event with a certain generated value of the kinematic variable to be reconstructed in a different bin, consist of twelve bins for the reconstructed spectrum and six bins for the generated distributions. Events with top quark transverse momentum beyond 240~GeV or absolute rapidity larger than 2.1 are shown in the overflow bins in Figs.~\ref{variable_top_pt},~\ref{variable_top_y} but are not considered in the unfolding.

\subsection{Estimation of systematic uncertainties}

Systematic uncertainties are estimated by repeating the fit and the unfolding procedure with templates affected by the systematic uncertainty and also, if the systematic uncertainty is affecting the signal process, variated migration matrices. The systematic uncertainties with the largest impact on the differential cross section measurement stem from the uncertainty on missing transverse energy, jet energy scale and the renormalization and factorization scale of the $t\bar{t}$ and W+heavy jets modeling.

\section{Results}

\begin{figure}[h]
\begin{minipage}{18pc}
\includegraphics[width=18pc]{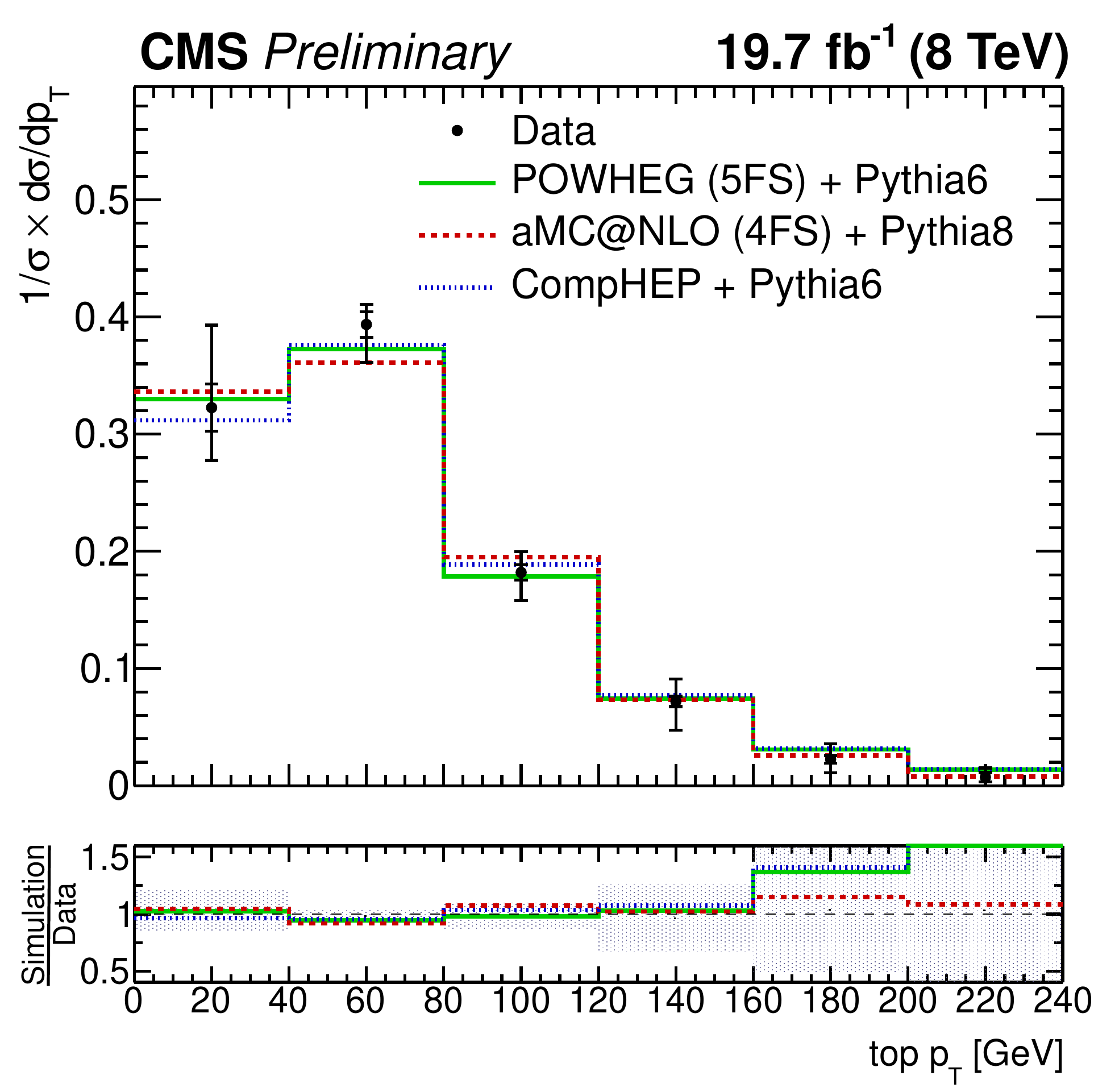}
\caption{\label{unfold_top_pt}Unfolded $p_{\mathrm{T}}$ distribution of the top quarks in the combined lepton+jets channel compared with predictions from \textsc{PowHeg+Pythia6} (solid),
\textsc{aMC@NLO+Pythia6} (dotted) and \textsc{CompHEP} (dashed). The inner error bars show the statistical uncertainty while the outer error bars show the full (stat.+syst) uncertainty.}
\end{minipage}\hspace{2pc}%
\begin{minipage}{18pc}
\includegraphics[width=18pc]{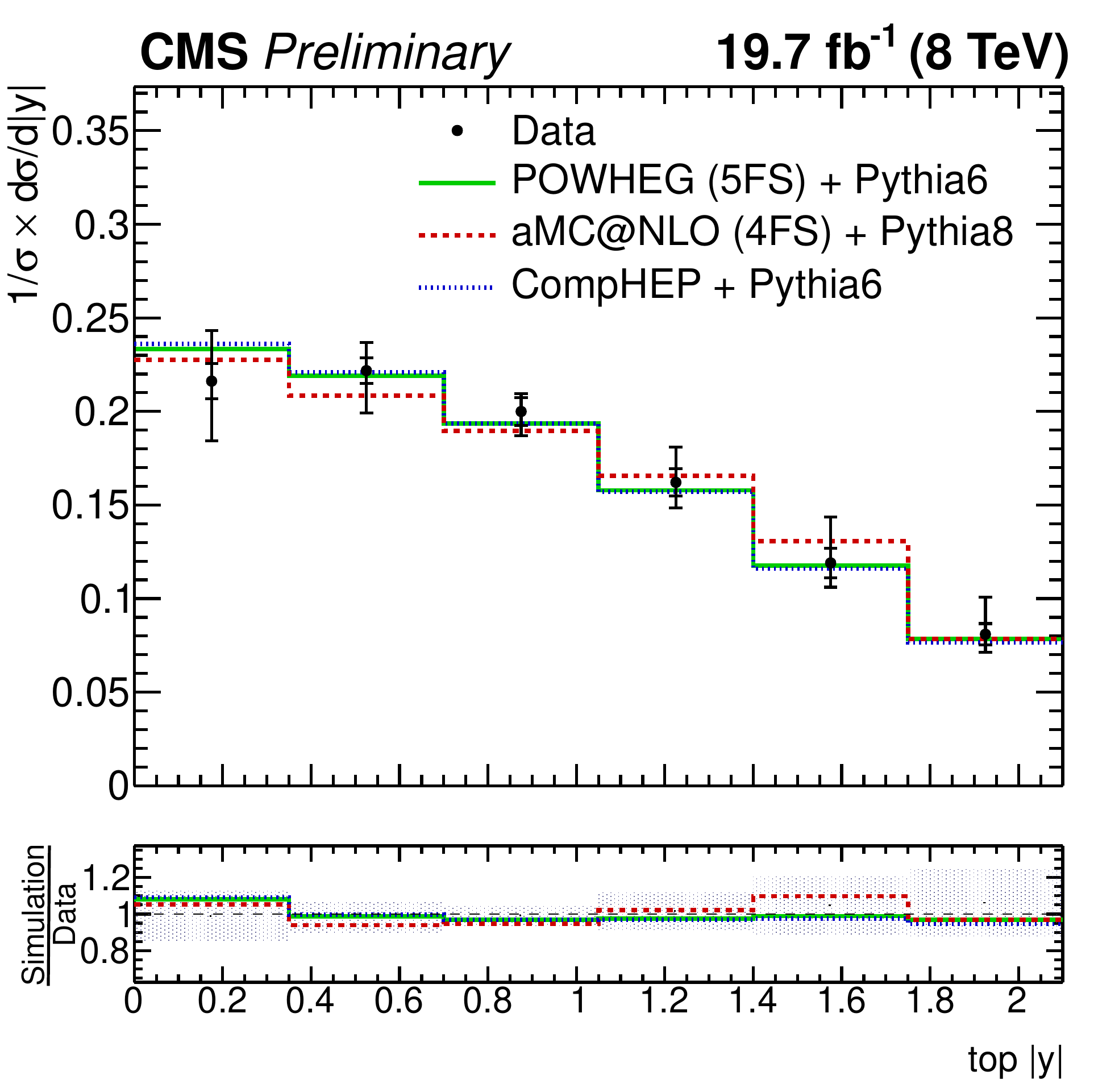}
\caption{\label{unfold_top_y}Unfolded $|y|$ distribution of the top quarks in the combined lepton+jets channel compared with predictions from \textsc{PowHeg+Pythia6} (solid),
\textsc{aMC@NLO+Pythia6} (dotted) and \textsc{CompHEP} (dashed). The inner error bars show the statistical uncertainty while the outer error bars show the full (stat.+syst) uncertainty.}
\end{minipage} 
\end{figure}

The unfolded distributions of the transverse momentum and the absolute value of the rapidity of the top quark, normalized to one by dividing by the measured inclusive cross sections, are shown in Fig.~\ref{unfold_top_pt} and Fig.~\ref{unfold_top_y}. The distributions from data are compared to generated distributions obtained with the generators \textsc{PowHeg}, \textsc{aMC@NLO} and \textsc{CompHEP}. While the two NLO generators differ in the used flavor scheme, \textsc{PowHeg} uses the 5FS and \textsc{aMC@NLO} the 4FS, the third generator \textsc{CompHEP} uses a matched sample consisting of LO $2\to2$ and $2\to3$ contributions. All three simulations describe the unfolded data within the estimated statistical and systematic uncertainties.

\section*{References}

\begin{thebibliography}{9}
\bibitem{Aaltonen:2009jj}
  CDF Collaboration,
  \textit{Phys.\ Rev.\ Lett.}\ {\bf 103} (2009) 092002
  [doi:10.1103/PhysRevLett.103.092002]
\bibitem{Abazov:2009ii}
  D0 Collaboration,
  \textit{Phys.\ Rev.\ Lett.}\ {\bf 103} (2009) 092001
  [doi:10.1103/PhysRevLett.103.092001]
\bibitem{Khachatryan:2014iya}
  CMS Collaboration,
  JHEP {\bf 06} (2014) 090
  [doi:10.1007/JHEP06(2014)090]
\bibitem{Aad:2014fwa}
  ATLAS Collaboration,
  \textit{CERN-PH-EP-2014-133} (2014)
  [arXiv:1406.7844 [hep-ex]]
\bibitem{Campbell:2009ss}
  J.~M.~Campbell, R.~Frederix, F.~Maltoni and F.~Tramontano,
  \textit{Phys.\ Rev.\ Lett.}\  {\bf 102} (2009) 182003
  [arXiv:0903.0005 [hep-ph]].
\bibitem{TOP-14-004}
  CMS Collaboration,
  CMS-TOP-PAS-14-004 (2014)
\bibitem{CMS}
  CMS Collaboration,
  \textit{JINST} {\bf 03} (2008) S08004
  [doi:10.1088/1748-0221/3/08/S08004]
\bibitem{TUnfold}
  S.~Schmitt,
  \textit{JINST} {\bf 7} (2012) T10003
  [doi:10.1088/1748-0221/7/10/T10003].
\end{thebibliography}
\end{document}